\documentclass[10pt,conference]{IEEEtran}
\usepackage{epsfig,graphicx,subfigure,psfrag,amsmath,cases,bm}
\usepackage{latexsym,amssymb,algorithm,mathtools}
\usepackage{algorithmic}
\usepackage{color}
\usepackage{scrtime}
\usepackage{stfloats}
\usepackage{tablefootnote}
\usepackage{cite}
\usepackage{amsfonts,mathabx}
\usepackage{textcomp}
\usepackage{caption}
\usepackage{xcolor,capt-of}
\usepackage{multirow}
\usepackage{subfigure}
 
\usepackage[left=0.625in,right=0.625in,bottom=0.95in,top=0.73in]{geometry}
\allowdisplaybreaks
\IEEEoverridecommandlockouts

\def\BibTeX{{\rm B\kern-.05em{\sc i\kern-.025em b}\kern-.08em
T\kern-.1667em\lower.7ex\hbox{E}\kern-.125emX}}

\begin{document}

\title{ Resource Allocation for UAV-Assisted Industrial IoT User with Finite Blocklength}

\author{%
\IEEEauthorblockN{Atefeh Rezaei\IEEEauthorrefmark{1}, Ata Khalili\IEEEauthorrefmark{2}, and Falko Dressler\IEEEauthorrefmark{1}}
\IEEEauthorblockA{\IEEEauthorrefmark{1} School of Electrical Engineering and Computer Science, TU Berlin, Germany}
\IEEEauthorblockA{\IEEEauthorrefmark{2} Institute for Digital Communications, Friedrich-Alexander-University Erlangen-Nurnberg, Germany}
\thanks{This work was supported by the Federal Ministry of Education and Research (BMBF, Germany) within the 6G Research and Innovation Cluster 6G-RIC under Grant 16KISK020K.}}

\maketitle
	
\begin{abstract}
We consider a relay system empowered by an unmanned aerial vehicle (UAV) that facilitates downlink information delivery while adhering to finite blocklength requirements. The setup involves a remote controller transmitting information to both a UAV and an industrial Internet of Things (IIoT) or remote device, employing the non-orthogonal multiple access (NOMA) technique in the first phase. Subsequently, the UAV decodes and forwards this information to the remote device in the second phase. Our primary objective is to minimize the decoding error probability (DEP) at the remote device, which is influenced by the DEP at the UAV. To achieve this goal, we optimize the blocklength, transmission power, and location of the UAV. However, the underlying problem is highly non-convex and generally intractable to be solved directly. To overcome this challenge, we adopt an alternative optimization (AO) approach and decompose the original problem into three sub-problems. This approach leads to a sub-optimal solution, which effectively mitigates the non-convexity issue. In our simulations, we compare the performance of our proposed algorithm with baseline schemes. The results reveal that the proposed framework outperforms the baseline schemes, demonstrating its superiority in achieving lower DEP at the remote device. Furthermore, the simulation results illustrate the rapid convergence of our proposed algorithm, indicating its efficiency and effectiveness in solving the optimization problem.
\end{abstract}
%

\section{Introduction}

Unmanned aerial vehicle (UAV)-aided wireless communication has drawn significant attention as a result of its simple deployment,  favorable channel gain, and more degrees of freedom \cite{Qingqing}. 
In some situations where the direct link is weak, the UAV can act as a relay to establish wireless connectivity between two or more nodes, increasing communication reliability and system throughput. \cite{Zeng1,Fan,Ata}. 
In \cite{Zeng1,Fan}, the authors considered a UAV as a relay to maximize the throughput by designing the source/UAV transmit power and the UAV's trajectory.
%
Recently, UAV-assisted non-orthogonal multiple access (NOMA) has been considered to improve spectrum efficiency and enhance connectivity through the flying UAV \cite{NOMA_UAV1,NOMA_UAV3}.
The authors in \cite{NOMA_UAV1} considered a max-min problem to optimize power allocation and UAV's altitude where a path-following algorithm was proposed.
In \cite{NOMA_UAV3}, a UAV-aided NOMA system was proposed where the time division multiple access (TDMA) and NOMA scheme were adopted to maximize the sum rate by jointly designing the NOMA precoding and trajectory of the UAV.

On the other hand, ultra-reliable low-latency communication (URLLC) services have been considered as one of the critical key performance indicators (KPIs) in the beyond fifth-generation (5G) networks that require low-latency transmission in applications such as healthcare, autonomous driving, tactical Internet, and remote control \cite{She}. 
However, meeting stringent latency and reliability requirements with a fixed infrastructure presents significant challenges \cite{TVTA,WPMC1}. To address this issue, UAVs are leveraged, benefiting from their inherent attributes such as flexible deployment, high maneuverability, and the ability to provide line-of-sight links.
In this regard, URLLC in the UAV network where the height of UAVs and the bandwidth were optimized to minimize the required total bandwidth for the URLLC service has been studied \cite{She}. 
In \cite{Pan}, researchers investigated a UAV-assisted relay network aiming to minimize the decoding error probability (DEP) by optimizing the blocklength assignment and the UAV's position. However, this study overlooked power allocation, which is a key aspect of resource allocation, and did not consider NOMA.
In \cite{ren}, they explored the concept of a UAV acting as an amplify-and-forward relay in a URLLC network, proposing a low-complexity iterative algorithm to minimize the packet DEP. Nevertheless, neither blocklength optimization for the robot nor cooperative NOMA was taken into account. 
Resource allocation of short packet transmission for mission-critical Internet of Things (IoT) networks to achieve URLLC under different schemes was studied in \cite{TWC_NC}.
In \cite{UAV_Two_Way}, the joint optimization of time, bandwidth, and UAV location in a two-way UAV relaying system was studied to maximize the average communication rate of the backward link while considering the URLLC requirement of the forward link.

Nonetheless, the studies mentioned above did not take into account the UAV Cooperative NOMA scheme for URLLC services which could further improve the performance gain of the system especially for the hidden users. In particular, reliability could be improved by exploiting NOMA transmission, as the entire blocklength could be used for transmission.   
However, exploiting NOMA for finite blocklength poses a major challenge as perfect successive interference cancellation (SIC) may not be guaranteed.
To address this knowledge gap,  we consider a cooperative network where the remote controller transmits the command message while adopting the NOMA scheme to the UAV and the remote device in the first phase. Subsequently, in the second phase, the UAV decodes and forwards (DF) the information of the remote device.
The direct path between the controller and the device is weak, making it challenging to meet strict latency and reliability requirements with fixed infrastructure.
Consequently, the UAV flies at a certain altitude to act as a relay and establish wireless communication between the controller and the remote device. 

Our main contributions can be summarized as follows:
\begin{itemize}
		
\item We optimize the DEP at the remote device while taking into account the UAV's DEP to find the NOMA power allocation coefficients, UAV's position, UAV's transmit power, and blocklength allocation between two links.

\item We propose an efficient solution based on the successive convex approximation to optimize DEP at the remote device which yields a suboptimal solution.
Simulation results show the effectiveness of the location optimization of the UAV and power allocation as compared to a fixed location and power allocation, respectively.
\end{itemize}

%

\section{System Model and Problem Formulation}

\begin{figure}
\centering
\includegraphics[width=1.01\linewidth]{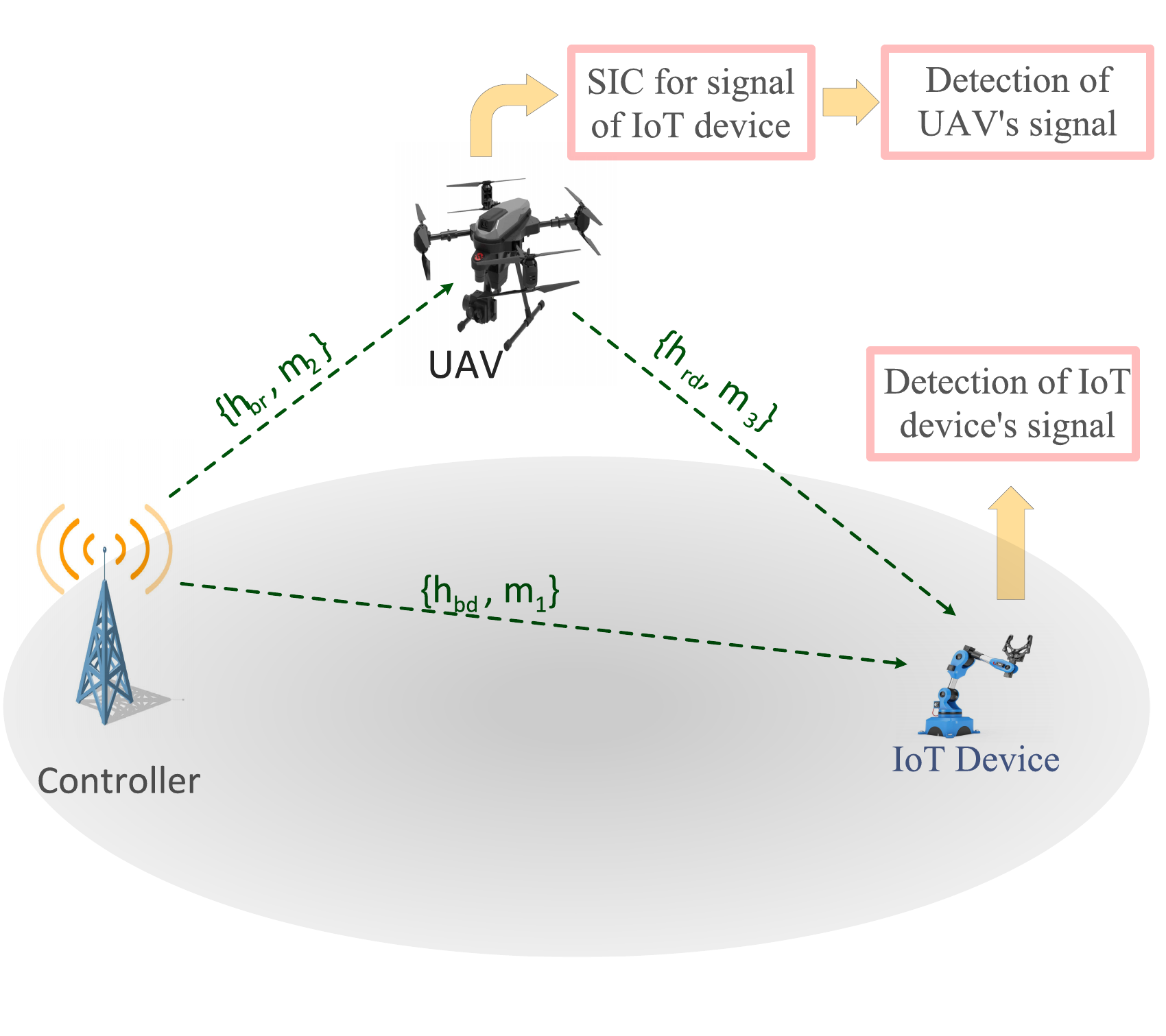}
\caption{UAV relay system with latency-constrained NOMA downlink.}
\label{fig:sys}
\vspace{-1em}
\end{figure}

In this paper, a downlink URLLC-enabled cooperative UAV-NOMA system is considered.
The system model consists of a controller, a UAV, and a remote device.
As shown in Fig.~\ref{fig:sys}, The remote controller is responsible for transmitting a command message to the distant remote device. To achieve further performance gains, a UAV is deployed between the controller and the remote device to assist in the transmission. Additionally, the remote controller needs to transmit two small packets, one to the UAV and the other to the remote device. Furthermore, we assume that all transceivers in the system are equipped with a single antenna. 
The location of the UAV, controller, and remote device are denoted by ${\mathbf{q}} = {[q_x,q_y,H]^{T}} \in {\mathbb{R}^{3 \times 1}}$, ${\mathbf{b}} = {[b_{x},b_{y} ,0]}^{T} \in {\mathbb{R}^{3 \times 1}}$, and ${\mathbf{d}} = {[d_{x}, d_{y} ,0]^{T}} \in {\mathbb{R}^{3 \times 1}}$, respectively, where $(\cdot)^T$ denotes the transpose operation.

The proposed cooperative network consists of two transmission phases. In the first phase, the UAV acts as a DF antenna relay, using the SIC technique to decode the information from the remote device. In the second phase, the UAV forwards the decoded information to its destination. The blocklengths assigned to the remote device and the UAV in the first phase are given by $m_1$ and $m_2$.
As a result of using the NOMA technology, we have $ m_1=m_2=m' $. 
The packet sizes for both the remote device and the UAV are assumed to be the same and are denoted as $D$ bits.
Also, the blocklengths assigned to the remote device in the second phase are denoted by $m_3$.

In URLLC services, the blocklength needs to be finite and short to ensure low-latency communication. To achieve this, a superposition coding (SC) technique is applied at the controller, enabling the transmission of signals from both the UAV and the remote device within the same resource block, but at different power levels.
The channel power gains between the controller-to-UAV and UAV-to-remote device are denoted by $h_{br}$ and $h_{rd}$, respectively.\\

Based on the free space channel model, we have
${h_{br}}= \frac{{{\beta _0}}}{{{{\left\| {{\mathbf{q}} - {\mathbf{b}}} \right\|}}}},$ and
${h_{rd}} = \frac{{{\beta _0}}}{{{{\left\| {{\mathbf{q}} - {\mathbf{d}}} \right\|}}}},$
where $\beta_{0}$ denotes the received power gain at the reference distance $d_{0} = 1$m. In contrast to the Shannon formula, the transmission data rate  with finite blocklength, can be approximated as \cite{Blocklength}
\begin{equation}\label{rate}
R_i \approx \log_2 (1 + {\gamma _i}) - {Q^{ - 1}}(\epsilon_i )\sqrt {\frac{1}{{m_i}}V_i},\ \forall i\in\{1,2,3\} ,
\end{equation}
where $\epsilon_i$ is the decoding error at the destination, $V_i$ is the channel dispersion, and $Q^{-1}(\cdot)$ is the inverse of the Gaussian Q-function.
Moreover, the channel dispersion $V_i$, $\forall i$ is given by $V_i=a^2\big(1-(1+\gamma_i)^{-2}\big)$, where $a=\log_2(e)$ and $\gamma_i$ is the signal-to-interference-plus-noise ratio (SINR) of the receiver $i$.
Indeed, in the proposed system, the controller encodes $R_i$ bits for receiver $i$ into a codeword with a blocklength of $m_i$ symbols, following the NOMA principle. Subsequently, during the first phase, the controller transmits the superposition of these two codewords toward both the UAV and the remote device.

In the following, we study the details of the transmission signals for the first and second phases.

\subsection{{First Phase: Direct Transmission}}

In this phase, the controller transmits a superimposed signal, including both information of the remote device and the UAV, by employing the power-domain NOMA scheme.
Thus, the transmit signal can be written as $s =\sqrt{p_{1}} s_{1} + \sqrt{p_{2}} s_{2}$, where ${s_1}$ and ${s_2}$ are transmitted symbols and are assumed to be independently circularly symmetric complex Gaussian (CSCG) distributed with zero mean and unit variance.
Furthermore, ${p_{1}}$ and ${p_{2}}$ represent the NOMA power allocation coefficients, which are required to satisfy the following constraint ${p_{1}}+ {p_{2}}\leq P_{\max}\label{power1}$.
Therefore, the received signal at the UAV and the remote device  can be expressed as
$y_{1} = {{h}_{br}}\bigg(\sqrt{p_{1}} s_{1} + \sqrt{p_{2}} s_{2}\bigg) + z_1,$~
$y_{2} = {{h}_{bd}}\bigg(\sqrt{p_{1}} s_{1} + \sqrt{p_{2}} s_{2}\bigg) + z_2,$
where $z_1 \sim \mathcal{N}(0,\,\sigma_{}^{2})\,$ and  $z_2 \sim \mathcal{N}(0,\,\sigma_{}^{2})\,$ are the received CSCG noise at the UAV and the remote device, respectively.

Then, the UAV applies the SIC technique to decode its own information. The UAV initially decodes the information of the remote device to obtain its own information successively. It is worth noting that the message of the remote device is encoded over all $m_2$ symbols, necessitating the UAV to collect all $m_2$ message-carrying symbols for successful decoding.
Without loss of generality, UAV can apply SIC to distinguish its own signal from $s_1$, whereas the remote device can only treat interference $s_2$ as noise.
Consequently, the received SINR at the  remote device in the first phase can be written as
\begin{equation}\label{SINR1.UAV}
\gamma _{1} = \frac{{p_{1}}|h_{bd}|^2}{{p_{2}}|h_{bd}|^2 + \sigma_{}^{2}},
\end{equation}
where $ h_{bd} $ shows the channel coefficient between the controller and the remote device following  Rayleigh fading distribution.
Also, the corresponding SNR of the UAV is given as 
\begin{equation}\label{SINR2.UAV}
\gamma _{2} = \frac{{p_{2}}|h_{br}|^2}{\sigma_{}^{2}}.
\end{equation}
We remark that to perform the SIC, the following constraint should satisfy
$|h_{br}|^{2}\geq |h_{bd}|^{2}$ as the remote device is far from the controller \cite{schober}.
	 
	 
\subsection{{Second Phase: Cooperative Transmission}}

In this phase, the UAV forwards the decoded information of the remote device.
If the UAV, successfully decodes the combined packet, it will forward the remote device's information with coding rate $\frac{D}{m_{3}}$. 
Thus, the received signal at the remote device in the second phase can be written as
$y_{3} =  \sqrt{p_u}{\rm{ }}{h_{rd}}{s_2} +  {z_3},$
where ${z_3} \sim \mathcal{N}(0,\,\delta_{}^2)$ is the received noise at the remote device, and $p_u$ is the transmit power of the UAV in the second phase.
Thus, the SNR at the remote device in the second phase can be expressed as
\begin{equation}\label{SINR phase 2.d}
\gamma_{3} = \frac{{p_u}{|h_{rd}|^2}}{\delta_{}^2} .
\end{equation}

\subsection{Decoding Error Analysis with Finite Blocklength}

In the finite blocklength scenario, compared to the ideal case, the DEP increases, resulting in an unavoidable decrease in the Shannon data rate.
We assume $ \bar{\epsilon}_1 $ and $ \bar{\epsilon}_2 $ as the effective error probability at the remote device and the UAV, respectively.
Also, $ \bar{\epsilon}_3 $ is the effective error probability at the remote device through the second phase.
The DEP is approximated by
$\epsilon\approx 
Q(f(\gamma,m,D))$,
where 
$f(\gamma,m,D)=\ln2\sqrt{\dfrac{m}{1-(1+\gamma)^{-2}}}(\log_2(1+\gamma_i)-\frac{D}{m})$.
It is worth mentioning that in the second phase, there is no interference, and thus the effective error probability is equal to $ \bar{\epsilon}_3=\epsilon_3 \approx Q(f(\gamma_3,m_3,\frac{D}{m_3})) $.
However, the effective error probability $ \bar{\epsilon}_2 $ depends on both the perfect and non-perfect SIC cases in the NOMA transmission.
Now, we define $ \mathbb{D}_{i}^{j}=0 $ for the case that the symbol $ s_i $ is correctly decode at the receiver $ j $ and $ \mathbb{D}_{i}^{j}=1 $ when the decoding is unsuccessful.
Thus, the DEP of $ s_1 $ at the UAV is 
$\epsilon_{12}=\mathcal{P}\{\mathbb{D}_{1}^{2}=1\}=Q(f{(\tilde{\gamma},m_1,\frac{D}{m_{1}})})$,
where
\begin{equation}\label{sic1}
\tilde{\gamma} _{} =\frac{{p_{1}}|h_{br}|^2}{{p_{2}}|h_{br}|^2 + \sigma_{}^{2}}.
\end{equation}

On the other hand, the effective DEP of $ s_2 $ at the UAV can be achieved using the marginal probability as
$\bar{\epsilon}_2=\mathcal{P}\{\mathbb{D}_{2}^{2}=1\}=\sum_{j=0}^{1} \mathcal{P}\{\mathbb{D}_{2}^{2}=1\mid\mathbb{D}_{1}^{2}=j\}\mathcal{P}\{\mathbb{D}_{1}^{2}=j\}$.

In the perfect SIC, the received SNR is $ \gamma_2 $ and, thus, $ \mathcal{P}\{\mathbb{D}_{2}^{2}=1\mid\mathbb{D}_{1}^{2}=0\}=\epsilon_2= Q(f(\gamma_2,m_2,\frac{D}{m_2}))$. 
Whereas, in the imperfect SIC, the SINR function includes the interference effect of $ s_1 $, which is given by  
\begin{equation}\label{sic_faild}
\gamma'=\frac{{p_{2}}|h_{br}|^2}{P_1|h_{br}|^2+\sigma_{}^{2}}.
\end{equation}

The DEP of $ s_2 $ conditioned on the failed SIC, i.e.,  $ \epsilon'_2=\mathcal{P}\{\mathbb{D}_{2}^{2}=1\mid\mathbb{D}_{1}^{2}=1\}$ is approximated by
\begin{align}\label{r_sic}
&\epsilon'_2  =
\left\{
  \begin{array}{ll}
    1, & R_2>\log_2(1+\gamma'), \\
    Q(f(\gamma',m_2,R_2)), & R_2\leq \log_2(1+\gamma').
  \end{array}
\right.
\end{align} 

The effective decoding error probability of $ s_2 $ at the UAV is described as
\begin{equation}\label{Error2}
\bar{\epsilon}_2=\epsilon_2(1-\epsilon_{12})+\epsilon'_2\epsilon_{12},
\end{equation}
where $\epsilon_{2}=Q(f(\gamma_{2},m_{2},\frac{D}{m_2}))$. 
In a similar manner, the DEP of the remote device decoding $s_{2}$ under the cooperative NOMA scheme can be written as
\begin{equation}
\bar{\epsilon_{1}}=\big(\epsilon_3(1-\epsilon_{12})+\epsilon_{12}\big)\epsilon_{1},
\end{equation} 
where $\epsilon_{1}$ is the DEP of the remote device when the remote device has to decode $s_{1}$ in the first phase and is given by $\epsilon_{1}=Q(f(\gamma_{1},m_{1},D))$.

\subsection{Optimization Problem}

We aim to minimize the effective DEP of the remote device by jointly optimizing the NOMA power allocation coefficients, the UAV's position, UAV's power allocation, and blocklength allocation.
Hence, the optimization problem can be mathematically formulated as
\begin{subequations}
		\begin{align}
		& \text{P1}: \mathop {{\rm{min}}} \limits_{\scriptstyle{m', m_3}, \atop{\scriptstyle{p _1},{p _2},p_{u},\mathbf{q}}}  \: \:\:\bar{\epsilon}_1\\
		\text{s.t.}~~
		&  {p _{1}} + {p _{2}} \leq P_{\max}, \label{power}\\
		& x_{\text{min}}\leq q_x \leq x_{\text{max}}\label{location}, \quad\quad\quad y_{\text{min}}\leq q_y \leq y_{\text{max}},\\
		&  m'+m_3\leq M,\label{Msize}\\
		&\bar{\epsilon_{2}}\leq \epsilon^{\max}_{2}\label{epsilonUAV}\\
		& m'(p_1+p_2)+p_um_{3}\leq E_{\text{tot}}\label{Energy}
		\end{align}
\end{subequations}
where \eqref{power} expresses the feasibility of the NOMA power allocation coefficients where $P_{\max}$ is the maximum transmit power of the remote controller.
Eq.\ (\ref{location}) imposes a limitation on the two-dimension position of the UAV where $x_{\text{min}}$ and $x_{\text{max}}$ are the minimum and maximum positions of the UAV on the X plane.
Similarly, $y_{\text{min}}$ and $y_{\text{max}}$ are the minimum and maximum locations of the UAV on the Y plane.
Eq. \eqref{Msize} depicts the relationship between the time slots considered in the transmission process, serving as a constraint that limits the latency at the receivers of both phases.
\eqref{epsilonUAV} ensures the DEP requirement for the UAV. 
Finally, Eq.\ \eqref{Energy} ensures the total energy consumption of the system is within a budget $E_{\text{tot}}$.

\smallskip
\textit{Remark 1:} It is evident that the objective function involves the multiplication of three terms, i.e., $\epsilon_{3}\epsilon_{12}\epsilon_{1}$, where this term becomes significantly smaller in comparison to the other terms. Therefore, we simplify the problem, leading to a new objective function that can be restated as $\bar{\epsilon_{1}}=\epsilon_{1}(\epsilon_{3}+\epsilon_{12})$.

\textit{Remark 2:} The function $\epsilon$ is neither convex nor concave with respect to either $m$ or $\gamma$. This is because the Q-function itself initially behaves as a concave function for negative values of its argument and then becomes convex. As a result, the overall behavior of $\epsilon$ is non-convex due to this combined behavior of the Q-function. To address this non-convexity, we make the assumption that the transmission data rate, i.e., $\frac{D}{m}$, is smaller than the Shannon capacity, i.e., $\log_2(1+\gamma)$. This ensures that the following conditions are held for the finite block length transmission: $\epsilon\leq \epsilon_{\max}=0.1$, $\gamma\geq 1$, and $\log_2(1+\gamma)\geq \frac{D}{m}$. Consequently, the error probability becomes convex with respect to blocklength and transmission power when $\gamma\geq \max\{ \frac{1}{5 \ln 2 \frac{D}{m}}, \frac{8}{45(\frac{D}{m})^{2}\ln^{2}2}\}$ \cite{Yao}.

%

\section{Solution to the Optimization Problem}

The optimization problem (P1) is non-convex, mainly due to the presence of coupling variables, subtractive terms in the rate functions, and non-convexity constraints \eqref{epsilonUAV} and \eqref{Energy}. To address this challenge, we propose an iterative solution based on the alternative optimization (AO) method, which provides a sub-optimal solution. Specifically, we decompose the main problem into three sub-problems, where we optimize the power allocation and blocklength in the first and second stages, respectively. Finally, we optimize the position of the UAV using the successive convex approximation (SCA) method.

\subsection{ Power Allocation Design}

In this stage, we consider a fixed position $\mathbf{q}$ and blocklengths $m'$ and $m_3$ to design the power allocation coefficients. To handle the non-convexity of \eqref{epsilonUAV}, we introduce three slack variables $\tilde{\mu}$, $\mu'$, and $\mu_1$, which serve as lower bounds for the SINR functions \eqref{SINR1.UAV}, \eqref{sic1}, and \eqref{sic_faild}, respectively. Additionally, we define the variable set $\boldsymbol{\Delta}P=\{p_1, p_2, \tilde{\mu},\mu',\mu_1, p_{u}\}$.
Consequently, the power allocation problem is described as: 
\begin{subequations}
			\begin{align}
			& \text{P}_2: \mathop {{\rm{min}}} \limits_{ {\scriptstyle{\boldsymbol{\Delta}_P}}}  \:\: \epsilon_{1}(\epsilon_{3}+\epsilon_{12})\\
			\text{s.t.}~~&  \eqref{power},~\eqref{epsilonUAV}-\eqref{Energy}\\
			& \frac{{p_{2}}|h_{br}|^2}{p_1|h_{br}|^2+\sigma_{}^{2}}\geq \mu',\label{mup}\\
			& \frac{{p_{1}}|h_{br}|^2}{{p_{2}}|h_{br}|^2 + \sigma_{}^{2}}\geq \tilde{\mu} _{},\label{mut}\\
			& \frac{{p_{1}}|h_{bd}|^2}{{p_{2}}|h_{bd}|^2 + \sigma_{}^{2}}\geq \mu_1,\label{muz}	
			\end{align}
\end{subequations}	 
where \eqref{mup}, \eqref{mut}, and \eqref{muz} are  non-convex functions of the transmission powers. 
Moreover, the objective function involves the multiplication of two $Q$ functions, which is also non-convex.
Let define $\bar{\epsilon}_{1}$ and $ \bar{\epsilon}_2 $ as
\begin{align}
	&\bar{\epsilon}_{1}=\epsilon_{1}(\varepsilon_{3})=\frac{1}{2}(\epsilon_{1}+\varepsilon_{3})^2-\frac{1}{2}(\epsilon_{1}^2+\varepsilon_{3}^{2}),\label{eps1}\\
           &\bar{\epsilon}_2 =\epsilon_2+\frac{1}{2}(\epsilon_{1 2}+\varepsilon_{2})^2-\frac{1}{2}(\epsilon_{1 2}^2+\varepsilon_{2}^2)\label{epsilon2a},
\end{align}
where $\varepsilon_{2}=\epsilon^{'}_2-\epsilon_2 $ and $\varepsilon_{3}=\epsilon_{3}+\epsilon_{12}$.
Equations \eqref{eps1} and \eqref{epsilon2a} are not convex because they involve the difference of two convex functions.
To tackle this issue, we adopt Taylor approximation on non-convex terms to make a globally lower bound as
\begin{align}
         & \tilde{\epsilon}_2 \triangleq \epsilon_2^{(t)}+\nabla^{T^{(t)}}_{\epsilon_2}(\textbf{v}-\textbf{v}^{(t)}),\label{taylor1}\\
         & \tilde{\epsilon}^2_{1 2} \triangleq\epsilon_{1 2}^{2^{(t)}}+\nabla^{T^{(t)}}_{\epsilon^2_{1 2}}(\textbf{v}-\textbf{v}^{(t)}),\label{taylor2}\\
         & \tilde{\epsilon}''^2_{2} \triangleq\epsilon_{	 2}''^{2^{(t)}}+\nabla^{T^{(t)}}_{\epsilon''^2_{ 2}}(\textbf{v}-\textbf{v}^{(t)}),\label{taylor3}\\
         &\tilde{\epsilon}''^2_{3} \triangleq\epsilon_{3}''^{2^{(t)}}+\nabla^{T^{(t)}}_{\epsilon''^2_{ 3}}(\textbf{v}-\textbf{v}^{(t)}),\label{taylor}\\
         & \tilde{\epsilon}_1^2 \triangleq\epsilon_1^{(t)^2}+\nabla^{T^{(t)}}_{\epsilon_1^2}(\textbf{v}-\textbf{v}^{(t)}).\label{taylor4}
\end{align}
Here, $(.)^{(t)}$ denotes the functions or variables at the previous iteration.
$ \nabla^{{(t)}}$ is the gradient vector regarding the corresponding variable set $\textbf{v}$, where in the first sub-problem $\textbf{v}=\{p_1, p_2, \tilde{\mu},\mu',\mu_1,p_u\}$.
Also,
$ \nabla^{{(t)}}_{\epsilon_2} $, $ \nabla^{{(t)}}_{\epsilon^2_{1 2}} $,  $ \nabla^{{(t)}}_{\epsilon{''^2}_{ 2}} $, $ \nabla^{{(t)}}_{\epsilon''^2_{ 3}} $, and $ \nabla^{{(t)}}_{\epsilon^2_{ 1}} $ are the gradients of $ \epsilon_2 $, $ \epsilon^2_{1 2} $, $ \epsilon''^2_{ 2} $,  $\epsilon''^2_{ 3}$, and $\epsilon^2_{1}$ at the previous iteration, respectively that are shown at the top of the next page.
Note that $ \dfrac{\partial \gamma_2}{\partial p_2}=\dfrac{|h_{br}|^2}{\sigma^2} $, $  \dfrac{\partial \gamma_3}{\partial p_u}= \dfrac{|h_{rd}|^2}{\delta^2}$, and $ \dfrac{\partial f_{\epsilon'_2}}{\partial \mu'}=0,     \:\forall \frac{D}{m_2} > \log_2(1+\mu')$ according to \eqref{r_sic}.
 Furthermore, gradient of $f{(\gamma_i,m,D)}$ based on the variable $ \gamma_i $ is described as
{\small\begin{equation}\label{g_f}\dfrac{\partial f}{\partial \gamma_i}=\dfrac{\ln_2 \sqrt{m_i}}{(1+\gamma_i)\sqrt{1-(1+\gamma_i)^{-2}}}\bigg( \dfrac{ \log_2(1+\gamma_i)-\frac{D}{m_i} }{(1+\gamma_i)^2\big(1-(1+\gamma_i)^{-2} \big)} +1
\bigg).
\end{equation}}

\begin{figure*}
\begin{align}\label{gradient_p}
 & \nabla_{\epsilon_2}=\bigg \langle \frac{\partial \epsilon_2}{\partial p_1}, \frac{\partial \epsilon_2}{\partial p_2} , \frac{\partial \epsilon_2}{\partial \tilde{\mu}}, \frac{\partial \epsilon_2}{\partial \mu'},  
 \frac{\partial \epsilon_2}{\partial \mu_1}, 
 \frac{\partial \epsilon_2}{\partial p_u}
 \bigg \rangle=
  \bigg \langle 0 ,  -\frac{1}{\sqrt{2\pi}}e^{-\frac{1}{2}f^2_{\epsilon_2}} \dfrac{\partial f_{\epsilon_2}}{\partial \gamma_2}\dfrac{\partial \gamma_2}{\partial p_2} , 0 , 0, 0 , 0 \bigg \rangle\\ 
 & \nabla_{\epsilon_{1 2}^2}= \bigg \langle 0 , 0 ,  -\frac{2\epsilon_{1 2}}{\sqrt{2\pi}}e^{-\frac{1}{2}f^2_{\epsilon_{1 2}}} \dfrac{\partial f_{\epsilon_{1 2}}}{\partial \tilde{\mu}} , 0 , 0 , 0 \bigg \rangle,~~ \nabla_{\epsilon_{1 }^2}= \bigg \langle 0 , 0 ,  0 , 0 , -\frac{2\epsilon_{1 }}{\sqrt{2\pi}}e^{-\frac{1}{2}f^2_{\epsilon_{1 }}} \dfrac{\partial f_{\epsilon_{1 }}}{\partial {\mu_1}} , 0 \bigg \rangle\\
 & \nabla_{\epsilon''^2_2}=\bigg  \langle 0, \frac{2(\epsilon'_2-\epsilon_2)}{\sqrt{2\pi}}e^{-\frac{1}{2}f^2_{\epsilon_2}} \dfrac{\partial f_{\epsilon_2}}{\partial \gamma_2}\dfrac{\partial \gamma_2}{\partial p_2}, 0, -\frac{2(\epsilon'_2-\epsilon_2)}{\sqrt{2\pi}}e^{-\frac{1}{2}f^2_{\epsilon'_2}} \dfrac{\partial f_{\epsilon'_2}}{\partial \mu'} , 0 , 0\bigg \rangle \\
 & \nabla_{\epsilon''^2_3}=\bigg  \langle  
 0, 0 , 
  -\frac{2(\epsilon_3+\epsilon_{12})}{\sqrt{2\pi}}e^{-\frac{1}{2}f^2_{\epsilon_{12}}} \dfrac{\partial f_{\epsilon_{12}}}{\partial \tilde{\mu}}
  , 0 , 0 , \dfrac{2(\epsilon_3+\epsilon_{12}) }{\sqrt{2\pi}}  e^{-\frac{1}{2}f^2_{\epsilon_3}} \dfrac{\partial f_{\epsilon_3}}{\partial \gamma_3}\dfrac{\partial \gamma_3}{\partial p_u}          
     \bigg \rangle ,
\end{align}
\hrule
\end{figure*}

    In the case of $ \dfrac{\partial f}{\partial \mu} $, \eqref{g_f} can be employed as $\mu$ function which is a lower bound definition of $\gamma$. 
To solve non-convex constraints \eqref{mup}, \eqref{mut}, and \eqref{muz}, we introduce slack variables $ \zeta' $, $ \tilde{\zeta} $, and $ \zeta'' $. By utilizing Taylor approximation, we can handle the non-convexity of these constraints, following the approach presented in \cite{Atefeh}
 \begin{subequations}
 			\begin{align}
 			& {p_{2}}|h_{br}|^2\geq  \mu'\zeta'=\frac{1}{2}(\mu'+\zeta')^2-\frac{1}{2}(\mu'^2+\zeta'^2)\nonumber\\
 			& \simeq \frac{1}{2}(\mu'+\zeta')^2-\frac{1}{2}((\mu'^2)^{(t)}+(\zeta'^2)^{(t)})\nonumber\\
 			& -\zeta'^{(t)}(\zeta'-\zeta'^{(t)})- \mu'^{(t)}(\mu'-\mu'^{(t)}) ,\label{z1}\\
	& {p_{1}}|h_{br}|^2\geq \tilde{ \mu}\tilde{\zeta} \simeq \frac{1}{2}(\tilde{\mu}+\tilde{\zeta})^2-\frac{1}{2}((\tilde{\mu}^2)^{(t)}+(\tilde{\zeta}^2)^{(t)})\nonumber\\
 	& -\tilde{\zeta}^{(t)}(\tilde{\zeta}-\tilde{\zeta}^{(t)})- \tilde{\mu}^{(t)}(\tilde{\mu}-\tilde{\mu}^{(t)}) ,\label{z2}\\
	&{p_{1}}|h_{bd}|^2\geq{ \mu_1}{\zeta''} \simeq \frac{1}{2}({\mu_1}+{\zeta''})^2-\frac{1}{2}(({\mu_1}^2)^{(t)}+({\zeta''}^2)^{(t)})\nonumber\\
 	& -{\zeta''}^{(t)}({\zeta''}-{\zeta''}^{(t)})- {\mu_1}^{(t)}({\mu_1}-{\mu_1}^{(t)}),\label{z3}
 			\end{align}
\end{subequations}
where 
\begin{subequations}
 			\begin{align}
 			& p_1|h_{br}|^2+\sigma_{}^{2}\leq \zeta',\label{zetap}\\
 			& {p_{2}}|h_{br}|^2 + \sigma_{}^{2}\leq \tilde{\zeta},\label{zetat}\\
			&{p_{2}}|h_{bd}|^2 + \sigma_{}^{2}\leq {\zeta''},\label{zetatt}
 			\end{align}
 \end{subequations}
By applying the SCA technique, the resource allocation problem is rewritten as
\begin{subequations}
			\begin{align}
			& \text{P}3: \mathop {{\rm{min}}} \limits_{ {\scriptstyle{\zeta', \tilde{\zeta}, \zeta'',~ \boldsymbol{\Delta}_P}}}  \:\: \frac{1}{2}(\epsilon_{1}+\varepsilon_{3})^2-\frac{1}{2}(\tilde{\epsilon}_{1}^2+\tilde{\epsilon}{''}_3^2)\\
			\text{s.t.}~~&  \eqref{power},~\eqref{Energy}, \eqref{z1}-\eqref{z3}, \eqref{zetap}-\eqref{zetatt}	,\nonumber\\
			& \epsilon_2+\frac{1}{2}(\epsilon_{1 2}+\epsilon{'}_2-\tilde{\epsilon}_2)^2-\frac{1}{2}(\tilde{\epsilon}_{1 2}^2+\tilde{\epsilon}{''}_2^2)\leq \epsilon^{\max}_{2}
			\end{align}
\end{subequations}	 
Now, this problem is convex and can be effectively solved by utilizing the well-known optimization toolbox in MATLAB, specifically the "fmincon" function. The "fmincon" function is tailored for addressing nonlinearly constrained optimization problems, making it a suitable choice for the current scenario.

	
	
   
\subsection{Blocklengths Design}    

In this stage, we focus on designing the blocklengths for the given power coefficients, namely $p_{1}$, $p_{2}$, and $p_{u}$. As the system model comprises two phases, a trade-off is established between $m'$ and $m_3$. Specifically, increasing the blocklength $m'$ results in a decrease in the blocklength $m_3$, and vice versa.
 Therefore, we optimize the blocklengths while considering the given power coefficients, in order to strike the right balance between the two phases.
Thus, $ \text{P}_1 $ can be reformulated as
\begin{subequations}
			\begin{align}
			& \text{P3}: \mathop {{\rm{min}}} \limits_{ {\scriptstyle{\Delta_M}}}  \:\:  \epsilon_{1}(\epsilon_{3}+\epsilon_{12}) \\
			\text{s.t.}~~&  \eqref{Msize} , \eqref{epsilonUAV}, \eqref{Energy}.
			\end{align}
\end{subequations}   
We remark that $ \text{P}3 $ is an integer problem and can be solved optimally via exhaustive search.
However, in order to reduce the complexity, we limit the bounds of the integer variables.
To facilitate the search algorithm, we rewrite the expression of $\bar{\epsilon}_2$ as  $\bar{\epsilon}_2=\epsilon_2(1-\epsilon_{12})+\epsilon'_2\epsilon_{12}\geq \epsilon_2$.
Besides, for the practical application, the decoding error at the UAV should be very small, e.g., much lower than $0.5$.
As a result, we obtain the following lower bound for $m_2=m_1=m'$
\begin{equation}
m^{\text{lb}}_{2}>\frac{D}{\log_2(1+\gamma_{2})}.
\end{equation}

Using the blocklength constraint, we obtain 
$m^{\text{up}}_{3}\leq M-m_{2}$.
In order to ensure correct packet delivery in the second phase to the remote device, its reliability must be ensured.
As a result, the following inequality should be satisfied:
\begin{equation}
m^{\text{lb}}_{3}>\frac{D}{\log_2(1+\gamma_{3})}.
\end{equation}
Furthermore, based on \eqref{Msize}, we can derive an upper bound for $m_{2}$ as:
\begin{equation}
m_2^{\text{up}}=M- m_3^{\text{lb}}.
\end{equation}
Consequently, we perform an exhaustive search to find the optimal solution for $\text{P}_3$, and its complexity is reduced by taking advantage of limitations over the search parameters.
\subsection{Location Optimization}    

In the final stage, we optimize the position of the UAV while keeping the power coefficients and blocklengths fixed. As a result, $\text{P1}$ can be restated as:
\begin{subequations}
		\begin{align}
		& \text{P4}: \mathop {{\rm{min}}} \limits_{\scriptstyle{\textbf{q}}}  \: \:\: \epsilon_{1}(\epsilon_{3}+\epsilon_{12})\\
	&	\text{s.t.}~~ \eqref{location}, \eqref{epsilonUAV}.	
		\end{align}
\end{subequations}
To solve the location optimization, let define two slack variables as 
$ S_{br}\geq{{{\left\| {{\mathbf{q}} - {\mathbf{b}}} \right\|^2}}},\quad S_{rd}\geq{{{\left\| {{\mathbf{q}} - {\mathbf{d}}} \right\|^2}}}$ \cite{khalili2023energy}.
The underlying problem is still non-convex.
To deal with this issue, the SCA technique can be employed in each iteration.
Thus, we apply first-order Taylor expansion to obtain a globally lower-bound.
At the $t$-th iteration, we have
\begin{align}
	&{\left\| {{\mathbf{q}} - {\mathbf{b}}} \right\|^2}\geq {\left\| {{\mathbf{q}}^{(t)} - {\mathbf{b}}} \right\|^2}+2({ {{\mathbf{q}}^{(t)} - {\mathbf{b}}}})^T({ {{\mathbf{q}} - {\mathbf{q}^{(t)}}}})\leq S_{br},\label{q1}\\
	&{\left\| {{\mathbf{q}} - {\mathbf{d}}} \right\|^2}\geq {\left\| {{\mathbf{q}}^{(t)} - {\mathbf{b}}} \right\|^2}+2({ {{\mathbf{q}}^{(t)} - {\mathbf{d}}}})^T({ {{\mathbf{q}} - {\mathbf{q}^{(t)}}}})\leq S_{rd}.\label{q2}
\end{align}

Since the objective function is non-linear non-convex, let us introduce a new variables set $ \boldsymbol{\Delta}_q =\{S_{br}, S_{rd}\} $. We can use the same structure of the power allocation stage via the SCA in \eqref{taylor1}-\eqref{taylor}, where $ \textbf{v}=  \boldsymbol{\Delta}_q$ and the corresponding gradients for this stage are described on the top of the next page in which
\begin{align}
&\dfrac{\partial \gamma_2}{\partial S_{br}}=-\dfrac{p_2\beta_{0}^2}{\sigma^2S_{br}^2},~~\dfrac{\partial \gamma_3}{\partial S_{rd}}=-\dfrac{p_u\beta_{0}^2}{\sigma^2S_{rd}^2},\\   &\dfrac{\partial \tilde{\gamma}}{\partial S_{br}}=-\dfrac{p_1\beta{_0}^2\sigma^2}{(p_2\beta{_0}^2+\sigma^2S_{br})^2} ,\\
&\dfrac{\partial {\gamma'}}{\partial S_{br}}=-\dfrac{p_2\beta{_0}^2\sigma^2}{(p_1\beta{_0}^2+\sigma^2S_{br})^2} .
\end{align}
\begin{figure*}
	\begin{align}\label{gradient_q}
	 & \nabla_{\epsilon_2}=\bigg \langle \frac{\partial \epsilon_2}{\partial S_{br}}, \frac{\partial \epsilon_2}{\partial S_{rd}} \bigg \rangle= \bigg \langle   -\frac{1}{\sqrt{2\pi}}e^{-\frac{1}{2}f^2_{\epsilon_2}} \dfrac{\partial f_{\epsilon_2}}{\partial \gamma_2}\dfrac{\partial \gamma_2}{\partial S_{br}}, 0\bigg \rangle,~~ \nabla_{\epsilon_{1 2}^2}= \bigg \langle   -\frac{2\epsilon_{1 2}}{\sqrt{2\pi}}e^{-\frac{1}{2}f^2_{\epsilon_{1 2}}} \dfrac{\partial f_{\epsilon_{1 2}}}{\partial \tilde{\gamma}}\dfrac{\partial \tilde{\gamma}}{\partial S_{br}}, 0\bigg \rangle\\
	 & \nabla_{\epsilon''^2_2}=\bigg  \langle  \frac{2(\epsilon'_2-\epsilon_2)}{\sqrt{2\pi}}e^{-\frac{1}{2}f^2_{\epsilon_2}} \dfrac{\partial f_{\epsilon_2}}{\partial \gamma_2}\dfrac{\partial \gamma_2}{\partial S_{br}}  -\frac{2(\epsilon'_2-\epsilon_2)}{\sqrt{2\pi}}e^{-\frac{1}{2}f^2_{\epsilon'_2}} \dfrac{\partial f_{\epsilon'_2}}{\partial \gamma'}\dfrac{\partial \gamma'}{\partial S_{br}},0\bigg \rangle ,\\
	 	 & \nabla_{\epsilon''^2_3}=\bigg  \langle  \frac{2(\epsilon_3+\epsilon_{12})}{\sqrt{2\pi}}e^{-\frac{1}{2}f^2_{{\epsilon_{12}}}} \dfrac{\partial f_{{\epsilon_{12}}}}{\partial \tilde{\gamma}}\dfrac{\partial \tilde{\gamma}}{\partial S_{br}}  , \frac{2(\epsilon_3+\epsilon_{12})}{\sqrt{2\pi}}e^{-\frac{1}{2}f^2_{\epsilon_3}} \dfrac{\partial f_{\epsilon_3}}{\partial \gamma_3}\dfrac{\partial \gamma_3}{\partial S_{rd}}\bigg \rangle ,
	\end{align}
	\hrule
\end{figure*}

Consequently, the problem is converted to a convex optimization problem as
\begin{subequations}
				\begin{align}
				& \text{P}5: \mathop {{\rm{min}}} \limits_{ {\scriptstyle{\boldsymbol{ \mathbf{q},\Delta}_q}}}  \:\: \frac{1}{2}(\epsilon_{1}+\varepsilon_{3})^2-\frac{1}{2}({\epsilon}_{1}^2+\tilde{\varepsilon}_{3}^2)\\
				&\text{s.t.}~~ \eqref{location},\eqref{q1},\eqref{q2}\nonumber\\
				& \epsilon_2+\frac{1}{2}(\epsilon_{1 2}+\epsilon{'}_2-\tilde{\epsilon}_2)^2-\frac{1}{2}(\tilde{\epsilon}_{1 2}^2+\tilde{\epsilon}{''}_2^2)\leq \epsilon^{\max}_{2}
		\end{align}
\end{subequations}
Note that in the objective function, $ \epsilon_1 $ is independent of the location of the trajectory.
	
%

\section{Simulation Results}

We evaluate the performance gain of the proposed scheme via simulation.
The system bandwidth is set as $B = 1$ MHz and the downlink transmission delay duration is  $ 100 \:\mu\text{sec}$ that guarantees the criterion of the URLLC user in the industrial case.
The  DEP at the UAV is set as $10^{-8}$.
Besides, the total blocklength and the packet size are set as $M=100$ and $D=100$ bits, respectively.
The other parameters are set as follows: $H=80$~m, $\sigma^{2}=-174$~dBm/Hz, and $\beta_{0}^{2}=-50$~dB, $E_{\text{tot}}=10$ Joule, $x_{\min}=30$~m, $x_{\max}=120$~m, $y_{\min}=30$~m, $y_{\max}=120$~m, and $P_{\max}=40$ dBm.
	
\begin{figure}
\vspace{-.5em}
	\centering	\includegraphics[width=0.48\textwidth]{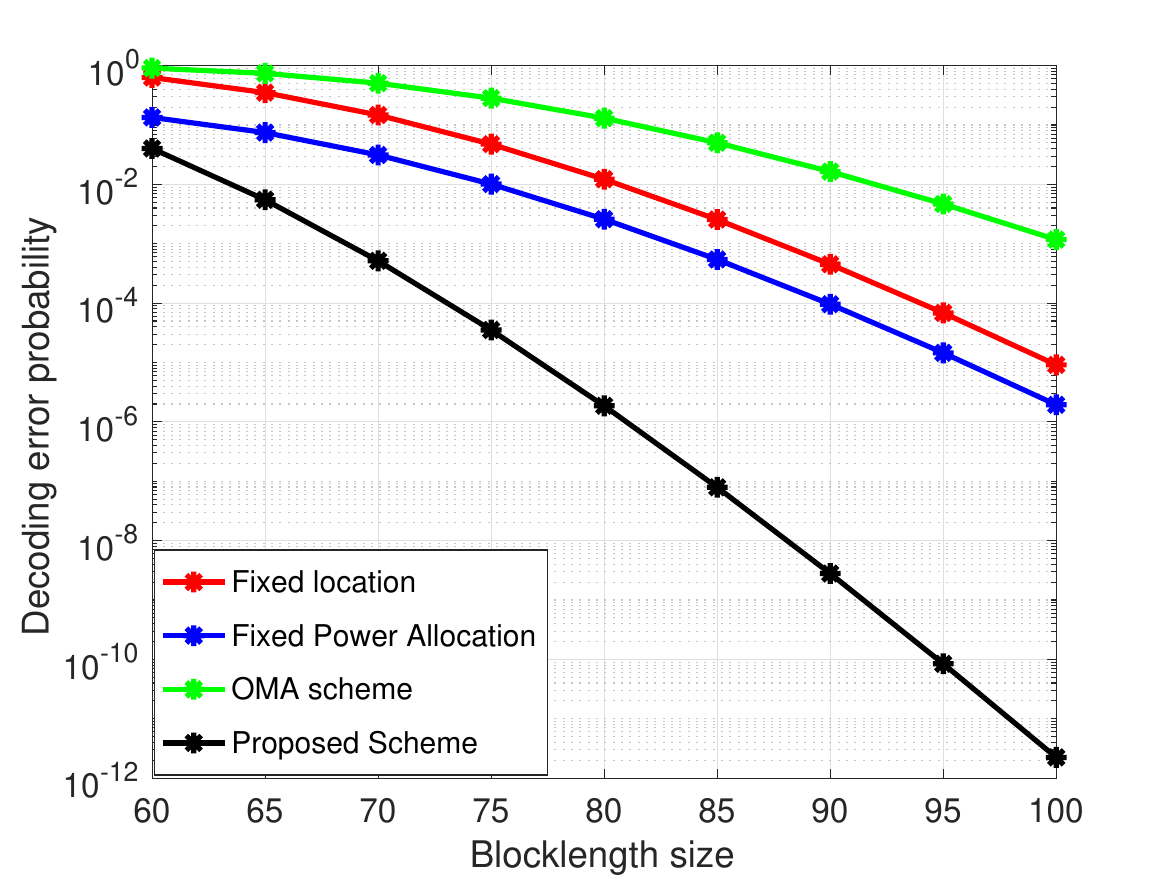}
	\caption{The DEP of the remote device versus different number of total blocklength.}
    \label{fig2}
\vspace{-.5em}
\end{figure}

Fig. \ref{fig2} demonstrates the impact of the total blocklength on the DEP at the remote device. Notably, as the packet length increases, the DEP at the remote device decreases, resulting in improved reliability.
Additionally, we present a comparison between our proposed algorithm and two other schemes: (1) the fixed location of the UAV, where the UAV is positioned in the middle between the remote controller and the remote device, and (2) the fixed power allocation scheme.
Remarkably, our proposed scheme outperforms both of these algorithms significantly in terms of DEP. This improved performance can be attributed to the optimized location of the UAV and the dynamic power allocation strategy employed in our proposed scheme.
These performance gaps demonstrate the validity of joint optimization.
Besides, we consider the OMA scheme in which the controller serves the remote device and the UAV in two different orthogonal channel uses or blocklengths. Hence, we formulate the DEP at the remote device in OMA-based optimization as
\begin{subequations}
		\begin{align}
		&  \mathop {{\rm{min}}} \limits_{\atop{\scriptstyle{p _1},{p _2},m_{1},m_{2}}}  \: \:\:\epsilon_1\\
		\text{s.t.}~~
		&\epsilon_{2}\leq \epsilon^{\max}_{2},~m_{1}p_{1}+m_{2}p_{2}\leq E_{\text{tot}}\label{EnergyOMA},\\
  &m_{1}+m_{2}\leq M,
		\end{align}
\end{subequations}

Fig. \ref{fig2} demonstrates that the cooperative NOMA scheme outperforms the OMA scheme in terms of performance. The reason behind this improvement lies in the UAV's ability to contribute and enhance the reliability of the remote device. By acting as a relay and strategically positioning itself, the UAV can significantly improve the received SNR at the remote device, resulting in higher reliability and better overall performance compared to the OMA scheme. Besides, this figure reveals that the NOMA scheme exhibits the best reliability performance. The superiority of the NOMA scheme can be attributed to the fact that the entire transmission blocklength can be dedicated to transmission, whereas in the OMA scheme, the blocklength needs to be divided into two parts for serving the UAV and the remote device separately. As a result, NOMA allows for more efficient and effective utilization of the blocklength, leading to enhanced reliability compared to OMA scheme.

\begin{figure}
	\centering
	\includegraphics[width=0.48\textwidth]{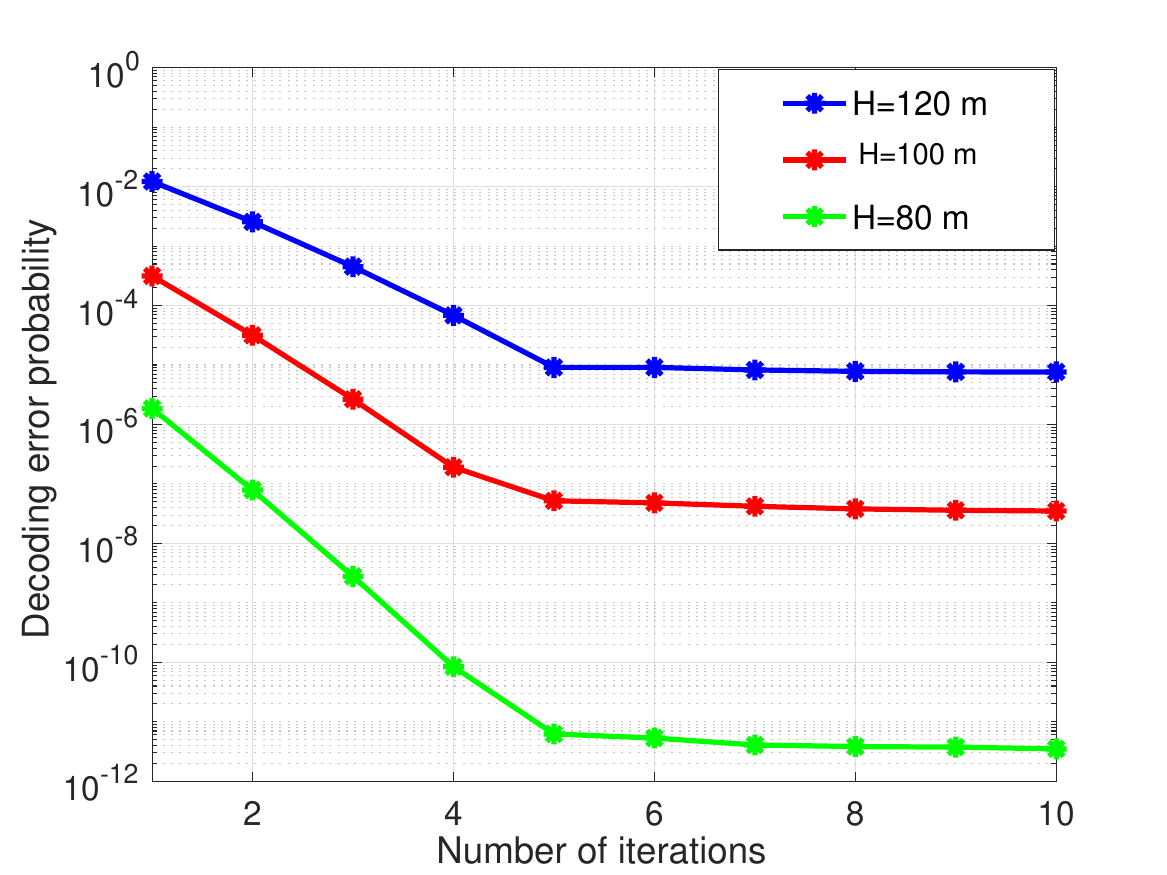}
	\caption{The DEP of the remote device versus the number of iterations for various UAV's height.}
    \label{fig3}
\vspace{-1em}
\end{figure}

In Fig. \ref{fig3}, we analyze the convergence behavior of our proposed algorithm for different heights of the UAV, denoted as $H$. The results highlight that our algorithm converges rapidly, as only 10 iterations are sufficient for convergence. Furthermore, the findings show that as the UAV's height ($H$) increases, the reliability decreases. This outcome can be attributed to the deteriorating quality of the channel with higher UAV heights, leading to a degradation of the SINR at the remote device. As the SINR at the remote device decreases, the DEP increases, resulting in lower reliability.

%

\section{Conclusion}
We investigated the problem of joint resource allocation and location optimization for cooperative UAV-NOMA networks to achieve low latency and high reliability supporting IIoT.
The problem of minimization of the DEP at the remote device was formulated while considering a low error probability threshold for the UAV.
The underlying problem was non-convex and an iterative low-complexity algorithm was proposed to obtain a sub-optimal solution.
Simulation results highlighted the significant performance gain achieved by employing the UAV as a relay in the cooperative UAV-NOMA system. This improvement was evident in terms of achieving lower DEP values, which directly translates to higher reliability and more efficient communication. In addition, the results clearly demonstrated the superiority of the cooperative NOMA scheme over OMA. This advantage arises from the fact that in the NOMA scheme, the entire transmission blocklength can be fully utilized for transmission. In contrast, the OMA scheme requires the blocklength to be divided into two separate parts to serve the UAV and the remote device independently.
\vspace{-2mm}
\bibliography{Mybib1}
\bibliographystyle{IEEEtran}

\end{document}